\documentstyle[aps,epsf,rotate,multicol]{revtex}

\newcommand{\be}{\begin{equation}}
\newcommand{\ee}{\end{equation}}

\begin{document}

\draft

\title{Self-Consistent Effective-Medium Approximations with Path Integrals}

\author{Yves-Patrick Pellegrini$^1$ and Marc Barth\'el\'emy$^{1,2}$}

\address{ $^1$ Service de Physique de la Mati\`ere Condens\'ee, \\
		Commissariat \`a l'Energie Atomique,\\
		BP12, 91680 Bruy\`eres-le-Ch\^atel, France.\\
	  $^2$ Center for Polymer Studies and Dept. of Physics,
		Boston University, Boston, MA 02215.}

\date{Last modified: December 9, 1999. Printed: \today\\
To be published in Physical Review E.}

\maketitle

\begin{abstract}
We study effective-medium approximations for linear composite media
by means of a path integral formalism with replicas. 
We show how to recover the Bruggeman
and Hori-Yonezawa effective-medium formulas. 
Using a replica-coupling ansatz, 
these formulas are extended into new ones
which have the same percolation thresholds as that of the Bethe 
lattice and Potts model of percolation, 
and critical exponents $s=0$
and $t=2$ in any space dimension $d\geq 2$. 
Like the Bruggeman and Hori-Yonezawa formulas, the new formulas 
are exact to second order in the weak-contrast and dilute limits. 
The dimensional range of validity of the four effective-medium 
formulas is discussed, and it is argued that the new ones are 
of better relevance than the classical ones in dimensions $d=3,4$ 
for systems obeying the Nodes-Links-Blobs picture, such as 
random-resistor networks.
\end{abstract}

\vskip 1cm


\section{Introduction}
Among various effective-medium formulas used to model the effective
behavior of random conducting linear composites, the symmetrical
Bruggeman formula \cite{BRUG35,LAND52} is undoubtly the most
popular. Applied to an insulator/conductor binary mixture, it predicts
a percolation-like transition \cite{KIRK71,CLER90,STAU92,SAHI98} for a
volumic fraction of conductor $p_c=1/d$, where $d$ is the space
dimension. The critical exponents are $s=t=1$, and its critical properties
are thoroughly discussed in Ref.\ \cite{CLER90}. This formula can be
interpreted in two ways. On one hand, Milton has shown \cite{MILT85} that it
yields the exact effective conductivity of an {\it ad hoc} ideal medium
built with a particular hierarchical structure. On the other hand, the
Bruggeman formula can be seen as a first (one-body) self-consistent
approximation to general disordered symmetric cell-materials
\cite{MILL69}, to which systematic corrections could be worked
out. However, the Bruggeman approximation is very different from a
mean-field theory of random conducting media. Indeed, an exact
mean-field calculation on the Bethe lattice \cite{STIN73,STRA74} predicts a
percolation threshold $p_c\sim 1/(2d)$ and exponents $s=0$ and
$t=3$. These exponents are exact for $d\geq 6$, as well as the
asymptotic behavior of the threshold when $d\to\infty$ (at least, for
the hypercubic lattice \cite{STAU92} to which a continuum theory
naturally compares \cite{NOTE1}). These values are also obtained in a
more systematic mean-field theory for random resistor networks
\cite{STEP77}.
The remarkable discrepancy between the mean-field results and
Bruggeman's formula indicates the ambiguous status of the Bruggeman
theory. As a matter of fact, in spite of various (mostly perturbative)
investigations \cite{HORI75,BERG81,LUCK91} in order to precise its
theoretical status, the reasons for the peculiar critical behavior of
Bruggeman's formula are not completely cleared up. 
More surprisingly, another
self-consistent effective-medium approximation \cite{HORI75,HORI77} due
to Hori and Yonezawa (HY), obtained for the same type of media by means
of a completely different approximation scheme (and later derived by
functional methods \cite{BART93}), exhibits the same exponents $s=t=1$
and a similar threshold behavior $p_c=1-\exp(-1/d)\sim 1/d$.

Apart from phenomelogical variants, and up to our knowledge,
the Bruggeman and HY effective-medium formulas are
the only ones obtained from the equations of
electrostatics in continuous media which are able to describe, at least
qualitatively, the overall features of a percolation transition in any
dimension. One intriguing question concerns the possibility of deriving
alternative effective-medium formulas from a continuum formulation, which
do not lead to the seemingly unavoidable values $s=t=1$ and $p_c\sim
1/d$. As we show in this paper, such a possibility exists. Our starting
point is the path integral approach recently put forward by
Barth\'el\'emy and Orland\cite{BART98}, where the effective-medium
problem is recast in a functional form. The problem reduces to
compute a free-energy: roughly, the logarithm, averaged over the
disorder, of a functional integral of Boltzmann-like weights, over
allowed field configurations (which include boundary conditions). The
average of the logarithm is carried out with the replica method (already
used in Ref.\ \cite{STEP77}). In Ref.\ \cite{BART98}, the authors showed
that the path integral formulation 
allows one to easily recover the second-order
weak-disorder expansion of the effective permittivity of nonlinear
composites \cite{BLUM91}.

However, this formulation
has not yet been used to derive self-consistent estimates.
In this paper, we show how this can be done. 
After a presentation of the functional approach to the homogeneization 
problem, and of the replica method (Sec.\ \ref{pifotp}), 
we discuss self-consistent effective-medium 
approximations (Sec.\ \ref{posa}). As usual
in such approximations, a background reference medium is introduced under the
form of an ansatz for the energy of the system, 
whose parameters are to be
determined self-consistently (Sec.\ \ref{o}). 
The new feature here is that the ansatz
contains a replica-coupling term, whose significance is
explained (Sec.\ \ref{rcacota}). 
The self-consistency conditions to determine its parameters 
are next discussed,
and two types of effective-medium formulas are identified (Sec.\ \ref{sc}):
one in which the replica couplings are cancelled
(hereafter referred to as ``type 1''), 
and the other one with non-zero replica couplings (``type 2'').
Two different approximations are then worked out for each type 
(Sec.\ \ref{tm}). It is found that 
type 1 generates the Bruggeman and HY formulas, 
whereas type 2 brings in two new effective-medium formulas which 
are ``replica coupling counterparts'' of the previous ones. 
They possess exponents $s=0$, $t=2$, and 
a threshold $p_c\sim 1/(2d)$ (Sec.\ \ref{nef}).
These new formulas are discussed in Sec.\ \ref{d}, 
where numerical results are presented
before we conclude in Sec.\ \ref{c}.

\section{Path integral formulation of the problem}
\label{pifotp}
The effective properties of a random 
conducting medium can be defined with the help of the
total dissipated power in the medium\cite{WILL86,SANC87,PONT92b}. 
In terms of the electric field
$E(x)$, the dissipated power $w$ in the system of volume $V$ reads
\begin{equation}
W[E]=\int_V dx\,w_x\bigl(E(x)\bigr), 
\end{equation}
where $w_x$ is the local power density.

{\em Hereafter, we take the volume $V$ of the sample equal to one}. In
heterogeneous materials, $w_x$ depends on constitutive
parameters randomly varying from point to point. For linear conducting
media with $j(x)=\sigma(x)E(x)$, where $\sigma$ is the local random
conductivity and $j$ is the electric current, we have
\begin{equation}
\label{linmed}
w_x\bigl(E(x)\bigr)=\sigma(x)E^2(x)/2.
\end{equation}
In the analogous effective
permittivity problem the dissipated power is replaced by the stored
energy $\varepsilon(x)E^2(x)/2$ ($\varepsilon$ is the
permittivity). For this reason, we shall abusively refer to $w_x$ as
the ``energy density'' hereafter. In the nonlinear problem, $w_x(E)$
is a non-quadratic function of $E$.

An alternative to solving Maxwell's equations is to minimize the total
energy $W$ subjected to the two constraints \cite{WILL86,SANC87}: (i) $E=- \nabla\phi$ and 
(ii) $\overline{E}=E_0$; here, the bar stands for a spatial average, 
and $E_0$ is a constant applied electric field. The 
minimum, $W^*\bigl(E_0\bigr)$, is expected to be self-averaging, as
occurs for the free-energy in disordered systems. We can therefore write
\begin{equation}
\label{mini}
W^*(E_0)=\left\langle\mathop{\text{min}}_{\overline{E}
=E_0 \atop{E=-\nabla\phi}} W[E]\right\rangle
\end{equation}
where the brackets $\langle\cdot\rangle$ denote the disorder average. 
$W^*(E_0)$ is the energy in a homogeneous medium 
characterized by an effective
constitutive law \cite{WILL86}
\begin{equation}
\label{constef}
\left\langle j\right\rangle={\partial W^*(E_0)\over \partial
E_0}=\sigma_{\text{eff}} E_0.
\end{equation}
The second equality defines the effective conductivity of the medium.

The problem thus reduces to computing the average of
the constrained minimum of a functional of the electric field. The
electric field derives from a potential and has a fixed mean value. We
can rewrite the constrained minimum in (\ref{mini}) using a path
integral
\begin{equation}
\mathop{\text{min}}_{\overline{E}=E_0\atop{E=-\nabla\phi}}
W[E]=-\lim_{\beta\rightarrow\infty}
\frac{1}{\beta}\mathop{\text{ln}}
\int{\cal D}E\,{\cal D}\phi\,
\delta(E+\nabla\phi)
\delta(\overline{E}-E_0)
e^{-\beta W[E]}.
\end{equation}
The minimum can be interpreted as the ground state energy
associated to the partition function
\be
\label{ZZ}
Z=\int \tilde{\cal D}E\,
e^{-\beta W[E]},
\end{equation}
where we have used the shorthand notation
\begin{equation}
\tilde{\cal D}E={\cal D}E\,\delta(\overline{E}-E_0)\int {\cal D}\phi\,
\delta(E+\nabla\phi) 
\end{equation}
for the constrained functional measure. 
We need to compute the average of the logarithm of (\ref{ZZ}).  In order
to proceed, we introduce replicas \cite{EDWA75,MEZA87} and use the identity
$\left\langle\ln Z\right\rangle=
\lim_{n\rightarrow 0}(\left\langle Z^n\right\rangle-1)/n$, hence
\begin{equation}
\label{wstar0}
W^*=-\lim_{\beta\rightarrow\infty}
\lim_{n\rightarrow 0}
\frac{1}{n\beta}
(\left\langle Z^n\right\rangle-1).
\end{equation}
The limits do not commute. The equivalent form
\begin{equation}
\label{wstar}
W^*=-\lim_{\beta\rightarrow\infty}\lim_{n\rightarrow
0}\frac{1}{n\beta}\ln\left\langle Z^n\right\rangle
\end{equation}
can be used as well. The replica method 
relies on the fact that one can easily compute
the replicated partition function 
$\langle Z^n\rangle$ for $n$ integer, and subsequently take the limit
$n\rightarrow 0$. The main quantity of interest therefore is
\be
\left\langle Z^n\right\rangle=
\int\prod_{\alpha=1}^{n}
\tilde{\cal D}E^{\alpha}\,\left\langle
e^{-\beta \sum_{\alpha=1}^{n} W[E^{\alpha}]}
\right\rangle.
\end{equation}
Denoting the replicated measure by 
$\tilde{\cal D}\bigl(E^\alpha\bigr)=\prod_{\alpha=1}^{n}
\tilde{\cal D}E^{\alpha}$, the average $\langle Z^n\rangle$ can be written 
in terms of an ``effective Hamiltonian''
\be
\label{zneff}
\langle Z^n\rangle=
\int\tilde{\cal D}\bigl(E^\alpha\bigr)\, e^{-\beta{\cal H}_e},
\ee
with
\begin{equation}
\label{heff}
{\cal H}_e=-{1\over\beta}\ln\left\langle e^{-\beta\sum_{\alpha=1}^{n}
W[E^{\alpha}]}
\right\rangle.
\end{equation}

For simplicity, we restrict ourselves to cell materials where the local
properties are statistically uncorrelated from site to site. Volume
integrals may then to be identified with sums over sites (each
pertaining to one cell) according to the correspondence $\int
dx\leftrightarrow v\sum_x$, where $v$ is an infinitesimal cell volume
(which defines the microscopic correlation length of the
problem). Then, ${\cal H}_e$ simplifies to
\begin{equation}
\label{calf}
{\cal H}_e=-{1\over\beta}\int\frac{dx}{v} \ln\left\langle 
e^{-\beta v\sum_\alpha w_x\bigl(E^\alpha(x)\bigr)}\right\rangle.
\end{equation}

Note that our discussion in Sec.\ III will be specialized to binary
disorder for which the constitutive parameters can take only two
values (but the proofs are general). That is, we assume that the
local energy density is distributed according to the probability
distribution
\be
P\bigl(w=w_x(E)\bigr)=p\delta \bigl(w-w_1(E)\bigr)+
q\delta \bigl(w-w_2(E)\bigr). 
\ee 
(where $q=1-p$). With this choice, 
\be
\label{znmoyen}
{\cal H}_e=-{1\over\beta}\int\frac{dx}{v} \ln
\left[p e^{-\beta v\sum_{\alpha=1}^n w_1\left(E^{\alpha}(x)\right)}
+qe^{-\beta v \sum_{\alpha=1}^n w_2\left(E^{\alpha}(x)\right)}\right].
\end{equation}
The above formalism applies to any form of the energy density, and in
particular to nonlinear media \cite{BERG92,PONT92a,YU94,PONT97}. 
A method for extracting from the path integral the second-order 
weak-contrast perturbation expansion of the effective potential 
$W^*(E_0)$, for nonlinear media, has been introduced in Ref.\ \cite{BART98}. 

\section{Principle of self-consistent approximations}
\label{posa}
In this paper, we consider the linear problem only. This section is devoted to self-consistent approximations to $W^*$. We first present
the principle for building such approximations through the 
introduction of a trial Hamiltonian. Then, we 
discuss the choice of a trial Hamiltonian with replica couplings. 
Finally, we explain how to exploit these replica couplings 
in order to obtain two kinds of self-consistent formulas.

\subsection{Overview}
\label{o}
 The common ingredient to the approximations discussed
below is the introduction of a linear comparison medium
described by a trial Hamiltonian ${\cal H}_0$ which is quadratic in the
electric field and non-random, e.g.\ the one-parameter ansatz
\begin{equation}
\label{trybrug}
{\cal H}_0=\frac{\sigma_0}{2}\int dx\,\sum_{\alpha}{E^{\alpha}}^2(x),
\end{equation}
where $\sigma_0>0$ is to be determined by an 
appropriate self-consistency condition. 
This Hamiltonian is that of a (replicated) homogeneous medium, 
but without couplings between replicas. Its meaning and that of 
other possible choices with replica couplings are discussed below.

The partition function $\langle Z^n\rangle$ can be rewritten as
\begin{equation}
\left\langle Z^n\right\rangle =
\frac{
\int{\tilde{\cal D}}\bigl(E^{\alpha}\bigr)\, e^{-\beta({\cal
H}_e-{\cal H}_0)}e^{-\beta{\cal H}_0}
}{
\int{\tilde{\cal D}}\bigl(E^{\alpha}\bigr)\,e^{-\beta{\cal H}_0}
}
\int{\tilde{\cal D}}\bigl(E^{\alpha}\bigr)\,e^{-\beta{\cal H}_0}
,
\end{equation}
or, with another notation
\begin{equation}
\label{znmoyen2}
\left\langle Z^n\right\rangle =
\left\langle
e^{-\beta({\cal H}_e-{\cal H}_0)}
\right\rangle_0 Z_0,
\end{equation}
where $Z_0$ is the partition function associated to ${\cal H}_0$, and
$\left\langle\cdot\right\rangle_0$ stands for the functional average with
weights $e^{-\beta{\cal H}_0}/Z_0$. Equ.\ (\ref{wstar}) thus reads
\begin{equation}
\label{wsplit}
W^*=W_0+\Delta W,
\end{equation}
where
\begin{eqnarray}
\label{wo}
W_0(E_0)&=&-\lim_{n\to 0\atop
\beta\to\infty}\frac{1}{n\beta}\ln Z_0,\\
\label{dwo}
\Delta W(E_0)&=&-\lim_{n\to 0\atop \beta\to\infty}\frac{1}{n\beta}
\ln \left\langle e^{-\beta({\cal H}_e-{\cal H}_0)}
\right\rangle_0.
\end{eqnarray}
The quantity $\Delta W(E_0)$ is difficult to compute (an exact evaluation
would lead to the exact result for the effective conductivity),
and we have to resort to approximations.

A natural self-consistency condition for ${\cal H}_0$ 
is 
\begin{equation}
\label{sccond}
\Delta W(E_0)=0,
\end{equation}
which completely determines ${\cal H}_0$ in the case where it 
depends on one single parameter, as in (\ref{trybrug}). For more general 
choices of ${\cal H}_0$ with several free parameters, (\ref{sccond}) 
only provides a relation between these parameters, and additional 
considerations are in order to determine them all. First of all, 
we have to precise the form of the ansatz to be used in our calculations.

\subsection{Replica couplings and choice of the ansatz}
\label{rcacota}
We deduce here the form of the trial Hamiltonian ${\cal H}_0$ from an
analysis of the effective Hamiltonian ${\cal H}_e$. Eq.\ (\ref{heff})
shows that the effective Hamiltonian is non-random but that the
average over disorder introduced a coupling between different
replicas. The meaning of these couplings is more transparent if we
carry out an expansion of (\ref{heff}) around the average field
$\overline{E}=E_0$ as in the weak-contrast expansion \cite{BART98}.
With $\partial_i=\partial/\partial E_i$ and $\Delta
E^\alpha=E^\alpha-E_0$ we have
\begin{eqnarray}
{\cal H}_e&=&n\langle w_x(E_0)\rangle\nonumber\\
&+&\frac{1}{2}
\left[ \sum_{\alpha}\int dx\,a_{ij}\Delta E^\alpha_i(x)
\Delta E^\alpha_j(x)-\beta\sum_{\alpha,\gamma}
\int dx dy\, c^{(2)}_{ij}(x-y)\Delta E^\alpha_i(x)
\Delta E^\gamma_j(y)\right] +\cdots,
\end{eqnarray}
where
\begin{eqnarray}
\label{aa}
a_{ij}&=&\langle \partial^2_{ij} w_x(E_0)\rangle,\\
\label{bb}
c^{(2)}_{ij}(x-y)&=&\langle\partial_{i} w_x(E_0)\partial_{j} w_y(E_0)
\rangle-\langle \partial_{i} w_x(E_0)\rangle\langle\partial_{j} 
w_y(E_0)\rangle.
\end{eqnarray}
The first non-zero replica-coupling term is proportional to $\beta
c^{(2)}$. We thus see that the coupling between replicas acts only
within clusters defined by $n$-point connected correlation functions
$c^{(n)}$, and accounts for the fluctuations of the electric field in
these clusters. The replica coupling would vanish if there were no
disorder at all. In the limit where the size of the region defined by
$c^{(2)}$ shrinks to zero -- which means that the system is observed
at a macroscopic level, we can approximate
\begin{equation}
c^{(2)}_{ij}(x-y)\simeq v c^{(2)}_{ij}(0)\delta(x-y),
\end{equation}
and we recover the expansion
\begin{equation}
\label{expan}
{\cal H}_e=n\langle w_x(E_0)\rangle+\frac{1}{2}\int dx\,\left[
\sum_{\alpha}a_{ij}\Delta E^\alpha_i(x)\Delta E^\alpha_j(x)-v\beta
\sum_{\alpha,\gamma}c^{(2)}_{ij}(0)\Delta E^\alpha_i(x)\Delta
E^\gamma_j(x)\right] +\cdots.
\end{equation}
which could directly be obtained from (\ref{calf}). The presence of
$v$ in front of the replica coupling term is the macroscopic remnant
of a microscopic average having been taken within a two-particle
cluster, of center $x$ and volume $v$. This discussion therefore
enlightens a relation between replica coupling and the electric field
fluctuations within clusters.

Expansion (\ref{expan}) suggests a two-parameter 
replica-symmetric ansatz of the form
\begin{equation}
\label{ansatz}
{\cal H}_0=\frac{1}{2} \sum_{\alpha\gamma}\int dx\, M^{\alpha\gamma} 
E^\alpha_i  E^\gamma_i,
\end{equation}
where
\begin{equation}
\label{matmok}
M^{\alpha\gamma}=\sigma_0\delta_{\alpha\gamma}-v\beta Q E_0^2.
\end{equation}
The free parameters are $\sigma_0$ and $Q$. Note that $Q$ has the
dimension of a squared conductivity, because it is related to a
quantity relative to two points. For simplicity, the ansatz $M$ is
diagonal in the euclidean vector space. However, we tried calculations
with a tensorial structure reproducing that of $a_{ij}$ and
$c^{(2)}_{ij}$ in Eqs.\ (\ref{aa}), (\ref{bb}); but, apart from a
different normalization for $Q$, no differences showed up in the final
effective-medium theories (as far as linear media are concerned).

An interesting feature of the ansatz (\ref{matmok}) is that, though 
being non-random, it embodies underlying disorder through its replica 
couplings. In order to understand this point, we compute 
$W_0(E_0)$ given by (\ref{wo})
\begin{equation}
\label{wopath}
W_0(E_0)=-\lim_{n\to 0\atop \beta\to\infty}\frac{1}{n\beta}\ln \int
\tilde{\cal D} E\,e^{-\frac{\beta}{2}\int
dx\,\left(\sigma_0\sum_\alpha{E^\alpha}^2-v\beta Q E_0^2
\sum_{\alpha,\gamma}{E^\alpha}\cdot{E^\gamma}\right)}.
\end{equation}
After writing $E=E_0-\nabla \phi$, and going to the Fourier transform 
of $\phi$ \cite{NOTE2}, we arrive at
\begin{eqnarray}
W_0(E_0)&=&-\lim_{n\to 0\atop \beta\to\infty}\frac{1}{n\beta}\ln
\left[(\mathop{\text{Det}}M)^{-1/2v}
e^{-\frac{\beta}{2}\sum_{\alpha\gamma}
M_{\alpha\gamma}E_0^2}\right]\nonumber\\
\label{wocomp}
&=&\frac{1}{2}\sigma_0(1-Q/\sigma_0^2) E_0^2.
\end{eqnarray}

Carrying out the derivative of (\ref{wopath}) 
with respect to $\sigma_0$, we obtain
\begin{equation}
\lim_{n\to 0\atop \beta\to\infty}\frac{1}{n}\left\langle\sum_\alpha
\langle{E^\alpha}^2\rangle\right\rangle_0=E_0^2+ \frac{Q}{\sigma_0^2}
E_0^2,
\end{equation}
where volume averages $\overline{E^2}$ have been replaced by statistical 
ones, the microscopic size $v^{1/d}$ being much smaller than 
that of the system, $V^{1/d}=1$.
All the replicas are equivalent, and the functional average $\langle\cdot\rangle_0$ selects in the limit $\beta\to\infty$ the real 
field in the medium. Hence, setting $\Delta E=E-E_0$, the previous 
equation leads to
\begin{equation}
\label{qq}
\frac{\langle\Delta E^2\rangle}{E_0^2}=\frac{Q}{\sigma_0^2}.
\end{equation}
which implies that $Q\geq 0$. When $Q\not=0$, the electric field
fluctuates in the medium, whereas it is uniform when $Q=0$. The ansatz
${\cal H}_0$ therefore represents a medium which is homogeneized
(because it is non-random), but which nonetheless
accounts for field fluctuations. 
We thus expect new effective medium approximations
when the replica coupling $Q$ is non zero.

\subsection{Self-consistency}
\label{sc}
Up to this point the discussion focused on the ansatz itself, without 
referring to ${\cal H}_e$. In particular, $\sigma_0$ was treated as a 
mere number. We now discuss what happens when self-consistency is 
imposed, within some approximation scheme. The medium is made of $N$ 
phases labelled by $\nu$, of respective conductivities $\sigma_\nu$ 
and volume concentrations $p_\nu$. The self-consistency relation 
$\Delta W(E_0)=0$, which imposes constraints on the ansatz, 
determines $Q$ as a function $Q=Q(\sigma_0,\{\sigma_\nu\})$. 
Then $W^*=W_0$ and, with (\ref{constef}),
\begin{equation}
\label{sigeff}
\sigma_{\text{eff}}=\sigma_0\left[1-\frac{Q(\sigma_0,\{\sigma_\nu\})}
{\sigma_0^2}\right].
\end{equation}

Suppose now that $\sigma_0=\sigma_0(\{\sigma_\nu\})$ is determined 
by an additional condition (to be precised below). Using the exact 
formula \cite{BERG78}  (cf.\ Appendix \ref{qfotf})
\begin{equation}
\label{berg}
\frac{\langle\Delta E^2\rangle}{E_0^2}=\sum_\nu \frac{\partial
\sigma_{\text{eff}}}{\partial \sigma_\nu} -1,
\end{equation}
the fluctuations of the electric field deduced from (\ref{sigeff}) 
can be written
\begin{equation}
\label{qberg}
\frac{\langle\Delta
E^2\rangle}{E_0^2}=\sigma_{\text{eff}}'(\sigma_0)\left(\sum_\nu
\frac{\partial \sigma_0}{\partial
\sigma_\nu}-1\right)+\frac{Q}{\sigma_0^2}-\frac{1}{\sigma_0}
\left[\frac{\partial Q}{\partial
\sigma_0}+\sum_\nu\left(\frac{\partial Q}{\partial
\sigma_\nu}\right)_{\sigma_0}\right]
\end{equation}
where the last derivative is performed at constant $\sigma_0$.  This
expression distinguishes between different contributions to the field
fluctuations: (i) the first term represents fluctuations coming from
the ``macroscopic'' background effective medium $\sigma_0$; (ii) the
second one is that already found in (\ref{qq}), and would be the only
one if $Q$ were independent from $\sigma_0$, and if 
$\sigma_0$ were equal to $\langle\sigma\rangle$,
 the trivial value corresponding
to a non-fluctuating reference medium for a multiphase composite, cf.\
Appendix \ref{qfotf}; (iii) and finally a third
term comes from the dependence of $Q$ on $\sigma_0$ and
$\sigma_\nu$. Both last terms are, according to the interpretation of
replica coupling developped in the previous section, of
``microscopic'' origin.

We now turn to the determination of $\sigma_0(\{\sigma_\nu\})$. A 
first obvious self-consistency condition for $\sigma_0$ is $Q\equiv 0$, 
so that $\sigma_{\text{eff}}=\sigma_0$. The effective-medium formulas obtained 
this way are referred to as ``type 1'' hereafter. As is shown 
below, to this type pertain the Bruggeman and HY formulas.

``Type 2'' effective-medium formulas are obtained by taking $\sigma_0$
as the solution of $\sigma_{\text{eff}}'(\sigma_0)=0$, and by using
this value in $\sigma_{\text{eff}}$. According to (\ref{qberg}), this
procedure makes the effective-medium insensitive to the fluctuations
generated in the reference medium $\sigma_0$, so that relevant fluctuations
only come from $Q$.

\section{Two approximations}
\label{tm}
In this section, the ideas introduced above are used within two
different approximations to $\Delta W(E_0)$, based on the ansatz
(\ref{ansatz}), (\ref{matmok}). For each approximation to $\Delta W$,
``type 1'' and ``type 2'' formulas are obtained.  Herafter,
$q=Q/\sigma_0^2$, so that
\begin{equation}
\label{sigefq}
\sigma_{\text{eff}}=\sigma_0(1-q).
\end{equation}

\subsection{One-impurity approximation}
\label{oia}

We first consider a ``one-impurity'' (or ``local'') calculation. The
Bruggeman formula emerges as the ``type 1'' effective-medium formula
in this approximation, which is not suprising since it can be seen as
a one-site (self-consistent) theory \cite{HORI75,LUCK91}.

\subsubsection{Approximation scheme}
The approximation for $\Delta W(E_0)$ [Eq.\ (\ref{dwo})], detailed in 
Appendix \ref{appb}, is a one-impurity approximation where interactions 
between different points are ignored. Let us denote by $w_0$ the 
trial Hamiltonian density, which depends on all the replicas, 
defined from (\ref{ansatz}), (\ref{matmok}) by
\begin{equation}
\label{hdens}
{\cal H}_0\equiv\int dx\, w_0[E(x)].
\end{equation}
Here and in Appendix \ref{appb}, the notation $[\cdot]$ indicates 
a dependence with respect to all the replicas.
Setting 
\begin{equation}
\label{dwx}
\Delta w_x[E(x)]=\sum_\alpha w_x\bigl(E^\alpha(x)\bigr)-w_0[E(x)],
\end{equation}
the one-impurity approximation results in
\begin{equation}
\label{oneimp}
\left\langle e^{-\beta({\cal H}_e-{\cal H}_0)}\right\rangle_0\simeq 
1+\frac{1}{v}\left\langle \left\langle e^{-\beta v \Delta w_x[E(x)]}\right\rangle_0-1\right\rangle.
\end{equation}
Because of statistical translation invariance, the final result is 
independant of the point $x$. The right-hand side can be computed 
exactly for any potential $w_x$ in the limit $\beta\to\infty$ using 
a saddle-point method. Setting $\Delta\sigma=\sigma-\sigma_0$ 
and
\begin{equation}
\mu=\left(1+\frac{\Delta\sigma}{d\sigma_0}\right)^{-1},
\end{equation}
we arrive at
\begin{equation}
\label{dwoi}
\Delta W(E_0)=\frac{1}{2}\langle\Delta\sigma \mu\rangle E_0^2
+\frac{q}{2}\langle \sigma\mu\rangle E_0^2.
\end{equation}
The condition $\Delta W(E_0)=0$ yields
\begin{equation}
\label{qbrug}
q=-\frac{\langle\Delta\sigma \mu\rangle}{\langle \sigma\mu\rangle}.
\end{equation}

\subsubsection{Type 1 formula: Bruggeman's}
Letting $q\equiv 0$ amounts to imposing $\langle\Delta\sigma 
\mu\rangle=0$, which is nothing but the Bruggeman equation
\begin{equation}
\left\langle\frac{\sigma-\sigma_0}{\sigma+(d-1)\sigma_0}\right\rangle=0.
\end{equation}
The Bruggeman equation can also be written $\langle\mu\rangle=1$, or $\sigma_0=\langle\sigma\mu\rangle$ if $d\not= 1$, or $\sigma_0=\langle\sigma\mu\rangle/\langle\mu\rangle$. 
The last expression is suitable for computing $\sigma_0$ iteratively 
(starting, e.g., from $\sigma_0=\langle\sigma\rangle$) in any 
dimension. The Bruggeman conductivity $\sigma_{\text{eff}}
=\sigma_0$ possesses a percolation threshold $p_c=1/d$, and critical 
exponents $s=t=1$ \cite{CLER90}.

The fluctuations computed from (\ref{berg}) read
\begin{equation}
\frac{\langle\Delta E^2\rangle}{E_0^2}=
\frac{\sigma_0\langle\mu^2\rangle}
{\langle\sigma\mu^2\rangle}.
\end{equation}

\subsubsection{Type 2 formula}
We now let $q\neq 0$ and given by (\ref{qbrug}) and
$\sigma_{\text{eff}}(\sigma_0)=\sigma_0\bigl(1-q(\sigma_0)\bigr)$. The
equation $\sigma_{\text{eff}}'(\sigma_0)=0$ reads
\begin{equation}
\label{eqsigo}
\sigma_0=\frac{\langle\sigma\mu\rangle}{\langle\mu\rangle}
\left(1+\frac{\langle\mu\rangle\langle\sigma^2\mu^2\rangle
-\langle\sigma\mu^2\rangle\langle\sigma\mu\rangle}
{2d\langle\sigma\mu\rangle^2}\right).
\end{equation}
Like Bruggeman's, this equation is easily solved by iterations starting
from $\sigma_0=\langle\sigma\rangle$. The iterations then always
converge to the physical solution, which we denote by
$\sigma_0^*$. The effective conductivity thus is
$\sigma_{\text{eff}}=\sigma_0^*\bigl(1-q(\sigma_0^*)\bigr)$.  To study
its critical behavior, we consider a binary mixture, where
$\sigma=\sigma_1$ with probability $(1-p)$, and $\sigma=\sigma_2$ with
probability $p$. In the conductor/superconductor limit where
$\sigma_2\to\infty$ we find, setting $p_c=1/(2d-1)$,
\begin{equation}
\label{sols0}
\sigma_0^*=\frac{\sigma_1}{p(d-1)}\left(\sqrt{\frac{1-p}{1-p/p_c}}
-1\right)\qquad (p<p_c)
\end{equation}
and
\begin{equation}
\label{solseff}
\sigma_{\text{eff}}=2\sigma_1\frac{\left[1-dp
-\sqrt{(1-p)(1-p/p_c)}\right]}{p^2(d-1)^2}\qquad  (p<p_c).
\end{equation}
The critical concentration $p_c$ can be interpreted as a percolation
threshold, and is the same as that obtained in the mean-field model on
a Bethe lattice\cite{STEP77} with connectivity $z=2d$. Since
$\sigma_{\text{eff}}=2(2d-1)\sigma_1/(d-1)\sim(p_c-p)^0$ for
$p\lesssim p_c$, the superconductivity exponent is $s=0$. Note however
that $\sigma_0$ displays a square-root cusp at $p=p_c$.  The critical
behavior for $p>p_c$ is obtained by exmining the insulator/conductor
mixture where $\sigma_2$ is finite and $\sigma_1=0$.  Then
\begin{eqnarray}
\sigma_0^*&=&\sigma_2\frac{p/p_c-1}{2(d-1)},\\
\sigma_{\text{eff}}&=&\sigma_2\frac{(p/p_c-1)^2}{4(d-1)^2 p} 
\qquad  (p>p_c).
\end{eqnarray}
Since $\sigma_{\text{eff}}\sim(p-p_c)^2$ for $p\gtrsim p_c$, 
the conductivity exponent is $t=2$.

For the special case of $d=1$, $\mu=\sigma_0/\sigma$ so that 
(\ref{eqsigo}) reduces to $\sigma_0=\langle1/\sigma\rangle^{-1}$, 
and $q=0$. Therefore, $\sigma_{\text{eff}}=\langle1/\sigma\rangle^{-1}$, 
which is the exact result. Like Bruggeman's, the new formula 
is also exact to second order in the 
contrast, in any dimension
\begin{equation}
\label{wcl}
\sigma_{\text{eff}}=\langle\sigma\rangle\left[1-\frac{\langle
\sigma^2\rangle-\langle\sigma\rangle^2}{d\langle\sigma\rangle^2}
+\cdots\right];
\end{equation}
and in the dilute limit where (e.g.) $p_2\ll 1$
\begin{equation}
\label{dl}
\sigma_{\text{eff}}=\sigma_1\left[1+d\frac{\sigma_2-\sigma_1}
{\sigma_2+(d-1)\sigma_1}+\cdots\right].
\end{equation}
In the discussion (Section \ref{d}), it is argued that because 
of its exponents $s\not= t$, and because it is less trivial than 
the Bruggeman formula (especially in the insulator/conductor case 
where it does not reduce to a straight line), this formula may 
constitute an easy-to-handle alternative to the latter in dimensions 
$d\geq 3$. Graphical comparisons between different effective-medium 
formulas are discussed in Sec.\ \ref{d}.

\subsection{Cumulant series approximation}
\label{asa}

In this section, we show how to recover by means of a cumulant
approximation to $\Delta W$ the effective-medium formula of HY,
together with its ``type 2'' counterpart.

\subsubsection{Approximation scheme}
We consider the first-order cumulant approximation
\begin{equation}
\label{cumul}
\left\langle
e^{-\beta({\cal H}_e-{\cal H}_0)}
\right\rangle_0\simeq
e^{-\beta\left\langle{\cal H}_e-{\cal H}_0\right\rangle_0}.
\end{equation}

We have then
\begin{equation}
\label{dwsa}
\Delta W(E_0)\simeq\lim_{n\to 0\atop \beta\to\infty}\frac{1}{n}
\left\langle
{\cal H}_e-{\cal H}_0\right\rangle_0.
\end{equation}
As is shown in Appendix C, the calculations here involve an expansion
in a series of the cumulants of the distribution of $\sigma$, whose
significance has been discussed at length in the original paper by
HY\cite{HORI77}. After some algebra, we obtain (cf.\ Appendix C)
\begin{equation}
\label{fmfodeu}
\Delta W(E_0)
=-{\sigma_0\over 2}\Bigl\{\left[1+d h_0
\bigl(1/(d\sigma_0)\bigr)\right]+d q h_0 
\bigl(1/(d\sigma_0)\bigr)\Bigr\}E_0^2,
\end{equation}
where
\begin{equation}
h_0(z)=\int_0^{\infty}du\, e^{-u}\ln\langle e^{-u\sigma z}\rangle.
\end{equation}
The family of functions $h_k$ is defined in Appendix C. 
The self-consistency $\Delta W(E_0)=0$ now yields
\begin{equation}
\label{q}
q=-\left[1+\frac{1}{d h_0
\bigl(1/(d\sigma_0)\bigr)}\right].
\end{equation}
  
\subsubsection{Type 1 formula: the Hori-Yonezawa formula}
\label{thfr}

We first consider the case with no couplings between replicas,
i.e. $q=0$. The HY formula for $\sigma_0$ reads
\begin{equation}
\label{hy}
h_0 \bigl(1/(d\sigma_0)\bigr)=-{1\over d},
\end{equation}
and the effective conductivity is $\sigma_{\text{eff}}=\sigma_0$. 
It can be shown that $\sigma_{\text{eff}}$ displays a percolation 
threshold
$p_c=1-\exp(-1/d)$, and exponents $s=t=1$\cite{HORI77}. 
Applying (\ref{berg})  and using (\ref{deriv}), the fluctuations read
\begin{equation}
\frac{\langle\Delta E^2\rangle}{E_0^2}=
-{2 +d h_1\bigl(1/(d\sigma_0)\bigr)\over 1+d h_1\bigl(1/(d\sigma_0)\bigr)}.
\end{equation}

\subsubsection{Type 2 formula}
\label{nef}

We now consider $q\neq 0$ and determined as a function of $\sigma_0$
by (\ref{q}). Then
\begin{equation}
\label{phiosol}
\sigma_{\text{eff}}(\sigma_0)=\sigma_0
\left[2+{1\over dh_0
\bigl(1/(d\sigma_0)\bigr)}
\right].
\end{equation}
The equation for $\sigma_0$ is $\sigma_{\text{eff}}'(\sigma_0)=0$; 
that is, with (\ref{deriv})
\begin{equation}
\label{deriveq}
2+{1\over d}{h_1\bigl(1/(d\sigma_0)\bigr)\over
h_0^2\bigl(1/(d\sigma_0)\bigr)}=0.
\end{equation}
In order to study the critical behavior of $\sigma_{\text{eff}}$, 
we consider again a binary mixture where $\sigma=\sigma_1$ with
probability $(1-p)$, and $\sigma=\sigma_2$ with probability
$p$. In the conductor/superconductor case where
$\sigma_2\to\infty$, we have
$h_0\bigl(1/(d\sigma_0))\bigr)=\ln(1-p)-\sigma_1/(d\sigma_0)$ and 
a similar equation for $h_1$, so that (\ref{deriveq}) reduces to 
a second-degree polynomial equation. Its physical solution reads
\begin{equation}
\label{solinf}
\sigma_0^*={2\sigma_1\over\sqrt{1+2d\ln(1-p)}
\left[1+\sqrt{1+2d\ln(1-p)}\right]}.
\end{equation}
It is defined for $p$ less than a critical value
\begin{equation}
p_c=1-e^{-1/(2d)}.
\end{equation}
This percolation threshold is the same as the one obtained in the Potts 
model at the mean-field level, and in the 
mean-field theory
of Ref.\ \cite{STEP77}. Reporting (\ref{solinf}) into (\ref{phiosol}), 
we arrive at
\begin{equation}
\sigma_{\text{eff}}={4\sigma_1\over
\left[1+\sqrt{1+2d\ln(1-p)}\right]^2}\qquad(p<p_c).\\
\end{equation}
Since $\sigma_{\text{eff}}\propto (p_c-p)^0$
for $p\lesssim p_c$, the superconductivity exponent is $s=0$.

In the opposite insulator/conductor case, where $\sigma_1=0$ 
and $\sigma_2$ is finite, the solution for $p>p_c$ can only be found perturbatively around the percolation threshold. Expanding the 
logarithm in $h_0$ and $h_1$ as
\begin{equation}
\ln\left[(1-p)+p
e^{-u\sigma_2/(d\sigma_0)}\right]=\ln(1-p)+\sum_{l\geq
1}{(-1)^{l-1}\over l}\left({p\over 1-p}\right)^l
e^{-lu\sigma_2/(d\sigma_0)},
\end{equation}
and defining
\begin{equation}
A(x)=\sum_{l\geq 1} {(-1)^{l-1}\over l^2} x^l=\int_0^x {dt\over t}
\ln(1+t),
\end{equation}
we find that 
\begin{equation}
\sigma_0^*={\sigma_2\over 4 d^2 A\bigl(p_c/(1-p_c)\bigr)}(p-p_c)+O
\left((p-p_c)^2\right),
\end{equation}
and that
\begin{equation}
\sigma_{\text{eff}}={\sigma_2\over
4d^2 A\bigl(p_c/(1-p_c)\bigr)}(p-p_c)^2+O\left((p-p_c)^3\right)
\qquad (p\gtrsim p_c),
\end{equation} 
where the conductivity exponent is $t=2$. Hence, as in the previous
``one-impurity'' approximation, the replica-coupling ansatz yields
critical exponents $s=0$, $t=2$, and an asymptotic dependence of the
threshold $p_c\sim 1/(2d)$ when $2d\gg 1$. One can easily check that
this new ``type 2'' effective-medium formula is exact to second order
in the weak-contrast limit and in the dilute limit.

After a few manipulations, we now obtain with (\ref{berg}) and (\ref{q}) 
\begin{equation}
{\left\langle \Delta E^2\right\rangle\over \left\langle E\right\rangle^2}
=2q\left(1+\frac{q}{2}\right).
\end{equation}

\section{Discussion}
\label{d}

\begin{figure}
\narrowtext
\vspace*{0.0cm}
\centerline{
\epsfysize=0.9\columnwidth{\rotate[r]{\epsfbox{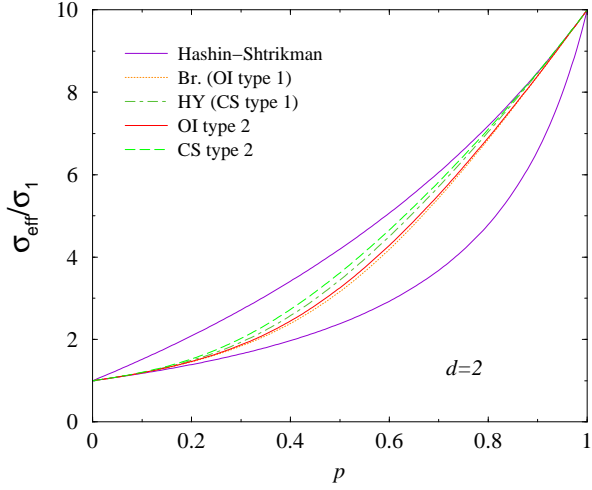}}}}
\vspace*{0.3cm}
\caption{Rescaled effective conductivities in dimension $d=2$ 
for a binary medium, versus the volume concentation $p$ of 
component 2. The conductivity ratio is $\sigma_2/\sigma_1=10$. 
Highest and lowest solid curves: Hashin-Shtrikman bounds; Br.: 
the Bruggeman formula (``type 1'', one-impurity 
approximation -- OI); HY: the Hori-Yonezawa formula (``type 1'',
cumulant series approximation -- CS); both ``type 2'' curves 
are the new formulas, within OI and CS approximations. 
}
\label{fig1} 
\end{figure}

\begin{figure}
\narrowtext
\vspace*{0.0cm}
\centerline{
\epsfysize=0.9\columnwidth{\rotate[r]{\epsfbox{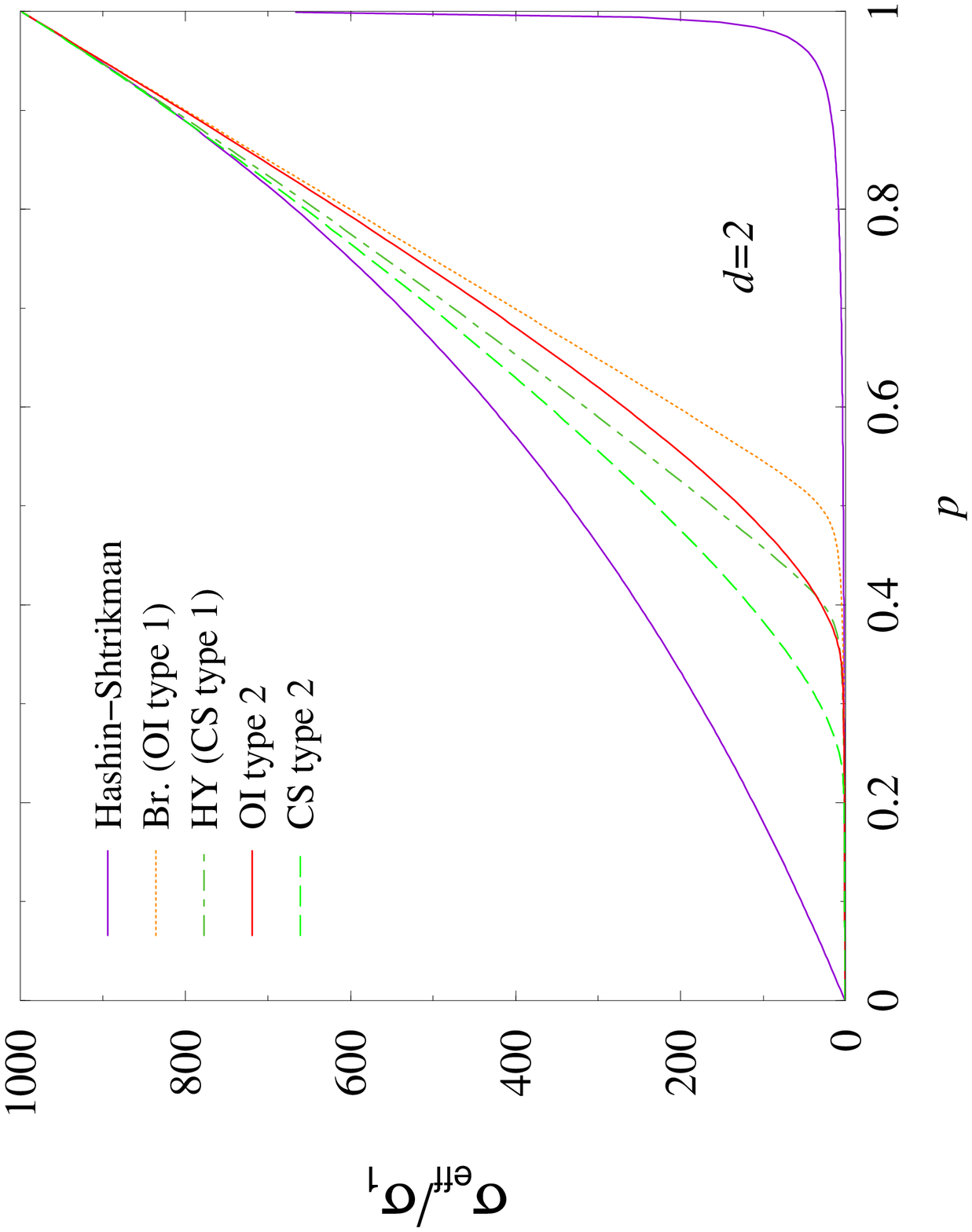}}}}
\vspace*{0.3cm}
\caption{Rescaled effective conductivities in dimension $d=2$ 
for a binary medium, versus the volume concentation $p$ of 
component 2. The conductivity ratio is $\sigma_2/\sigma_1=1000$.
Same plots as in Fig.\ \ref{fig1}.
}
\label{fig2}
\end{figure}

\begin{figure}
\narrowtext
\vspace*{0.0cm}
\centerline{
\epsfysize=0.9\columnwidth{\rotate[r]{\epsfbox{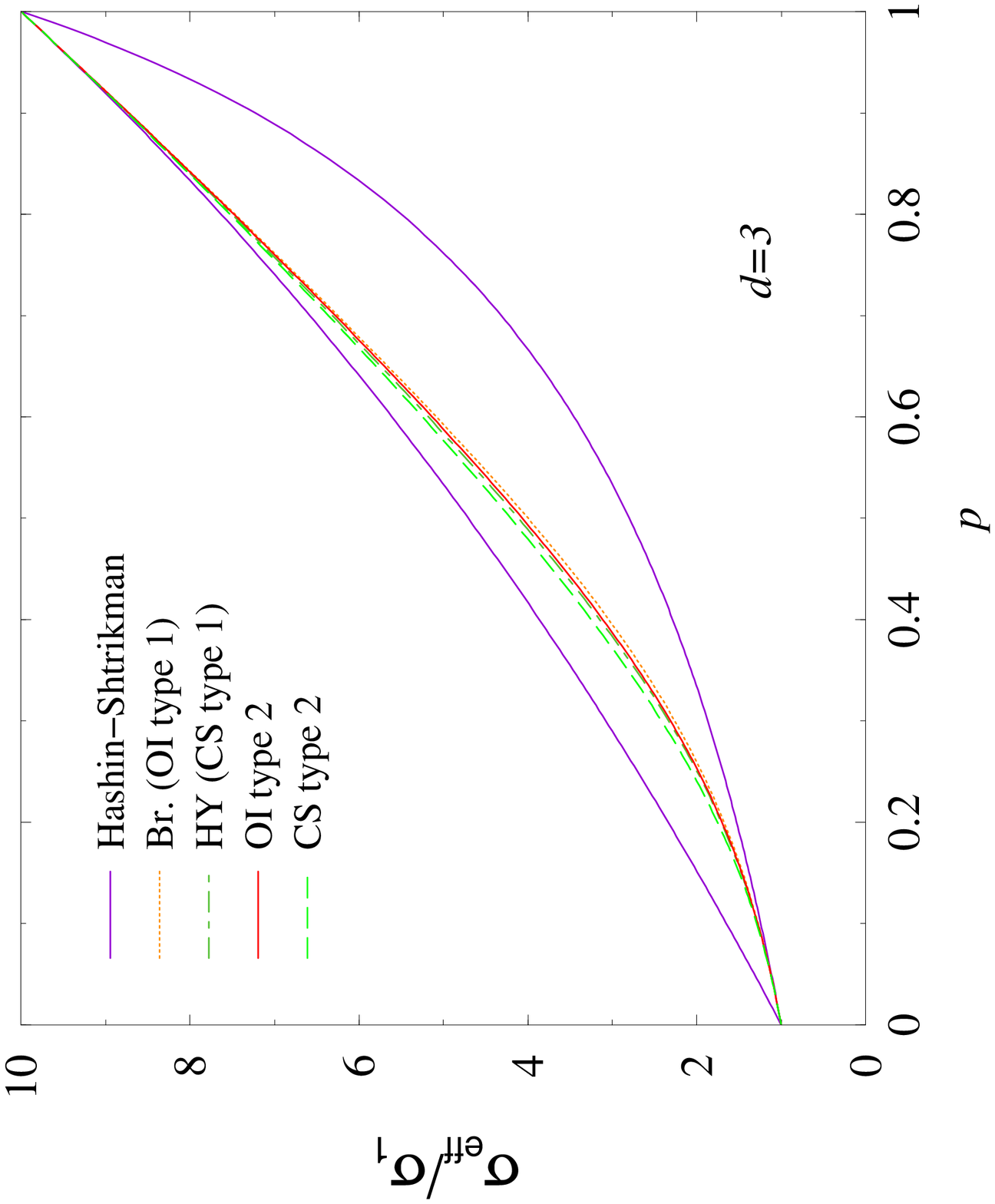}}}}
\vspace*{0.3cm}
\caption{Rescaled effective conductivities in dimension $d=3$ 
for a binary medium, versus the volume concentation $p$ of 
component 2. The conductivity ratio is $\sigma_2/\sigma_1=10$.
Same plots as in Fig.\ \ref{fig1}.\\
\\
\\
\\
}
\label{fig3}
\end{figure}

\begin{figure}
\narrowtext
\vspace*{0.0cm}
\centerline{
\epsfysize=0.9\columnwidth{\rotate[r]{\epsfbox{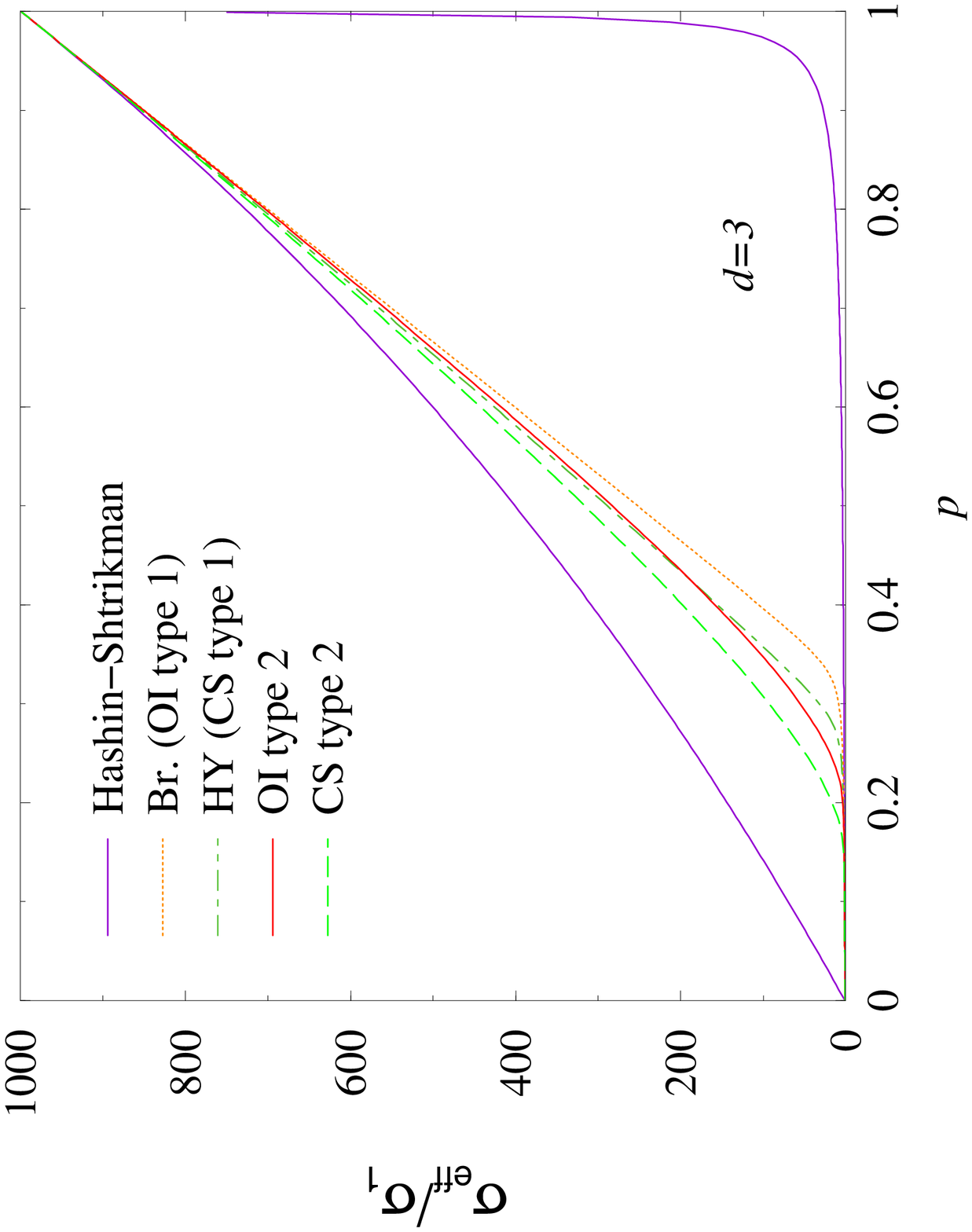}}}}
\vspace*{0.3cm}
\caption{Rescaled effective conductivities in dimension $d=3$ 
for a binary medium, versus the volume concentation $p$ of 
component 2.The conductivity ratio is $\sigma_2/\sigma_1=1000$.
Same plots as in Fig.\ \ref{fig1}.
}
\label{fig4}
\end{figure}

We plot in Fig. \ref{fig1} (resp.\ \ref{fig2}) the ``type 2'' scaled
conductivities $\sigma_{\text{eff}}/\sigma_1$ versus $p$, the volume
fraction of material 2, for a dielectric ratio $\sigma_2/\sigma_1=10$
(resp.\ $\sigma_2/\sigma_1=1000$). We also show the Hashin-Shtrikman
(HS) bounds \cite{HASH62}, the Hori-Yonezawa formula which comes from
a cumulant series (CS) approximation, and the one-impurity (OI)
Bruggeman formula.  The dimension is $d=2$. Figs.\ \ref{fig3} and
\ref{fig4} display similar plots for $d=3$.

For moderate contrast (Figs.\ 1 and 3), we observe that all four
self-consistent formulas lie close to each other. This is a
consequence of the fact that they are exact to second order in the
contrast. Also, for any contrast, the slopes at $p=0$ and $p=1$ are
all identical, which is a consequence of the fact that they are exact
to second order in the dilute limit $p\to 0$ (the expression near
$p=1$ is obtained by replacing $p$ by $p-1$ and by interchanging
$\sigma_1$ and $\sigma_2$). We also observe that the HS bounds are
satisfied in each case considered. However, the formulas obtained via
the cumulant series summation, i.e.\ both the HY formula and its
``type 2'' counterpart, do not reduce to the exact result
$\sigma_{\text{eff}}=\langle 1/\sigma\rangle^{-1}$ in dimension 1 (not
shown). This exact result is also the common value of the HS bounds
for $d=1$. Hence, formulas derived from the cumulant series
approximation do not obey the HS bounds in dimension $d=1$. On the
other hand, both the Bruggeman formula and its ``type 2'' counterpart
do reduce to the exact result when $d=1$, and can be seen to always
obey the HS bounds whatever $d$ is. The one-impurity approximation
scheme therefore appears to be of better physical relevance for all
dimensions, than the cumulant series approximation.

\begin{figure}
\narrowtext
\vspace*{0.0cm}
\centerline{
\epsfysize=0.9\columnwidth{\rotate[r]{\epsfbox{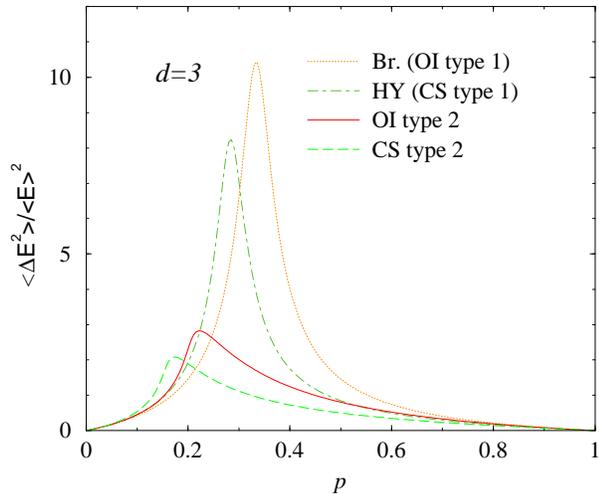}}}}
\vspace*{0.3cm}
\caption{Relative quadratic fluctuations of the field in dimension
$d=3$ for a binary medium, versus the volume concentation $p$ of
component 2. The conductivities are $\sigma_1=1$, $\sigma_2=1000$.
}
\label{fig5}
\end{figure}

We now discuss the critical behavior. First of all, the percolation
thresholds found in the``type 2'' formulas are
$p_c=1-\exp\bigl(-1/(2d)\bigr)$ (cumulant series) and $p_c=1/(2d-1)$
(one-impurity). These thresholds are the percolation thresholds of the
Potts model, and that of the Bethe lattice model, respectively. Both
thresholds decrease as $1/(2d)$ when $d\to\infty$, which is the exact
asymptotics. The Bethe and Potts models are mean-field models, where
emphasis is put on fluctuations in the couplings between a given site
and its neighbours. On the contrary, in effective-medium theories,
interactions between impurities are taked into account through the
self-consistent background medium. Such interactions are more
important for low dimensions. Effective-medium theories therefore
overestimate interactions in high dimensions, whereas mean-field
models are expected to underestimate them in low dimensions. Above the
upper critical dimension where mean-field models are accurate,
interactions between impurities become irrelevant. According to this
discussion, our ``type 2'' formulas appear as hybrids between
mean-field and usual effective-medium theories, and are expected to be
mostly relevant in dimensions intermediate between $d=1$ and the upper
critical dimension $d=6$.  Indeed, the condition
$\sigma'_{\text{eff}}(\sigma_0)=0$ minimizes the influence of the
background medium and, according to the interpretation developped in
Sec.\ \ref{rcacota}, replica coupling has to do with couplings
between neighboring points. The reason for which the introduction of a
replica-coupling ansatz yields the exact thresholds of mean-field
theories will have to be clarified in the future.  In Fig.\
\ref{fig5}, we plot the quadratic fluctuations $\left\langle\Delta
E^2\right\rangle/\left\langle E\right\rangle^2$ as a function of
$p$. The fluctuations in the ``type 2'' estimates are greatly reduced
compared to those of the Bruggeman and HY formula. This is consistent
with the fact that the influence of the background is reduced.

``Type 1'' formulas give exponents $s=t=1$, while for ``type 2''
formulas they are $s=0$, $t=2$. Mean-field theories yield $s=0$, $t=3$
which are the exact values for $d\geq 6$.  It is interesting to
compare these values to exact bounds deduced from the
Nodes-Links-Blobs (NLB) model, in all dimensions. The NLB model is
currently accepted as a good one for the backbone structure of real
random resistor networks\cite{STAU92}. The bounds read
\begin{mathletters}
\label{expbounds}
\begin{eqnarray}
&&t\ge 1+(d-2)\nu,\\
&&s\ge 1+(2-d)\nu,
\end{eqnarray}
\end{mathletters}
where $\nu>0$ is the correlation length exponent: $\xi\propto
|p-p_c|^{-\nu}$. They hold for $2\leq d\leq 6$, whereas for $d>6$ the
right-hand sides in (\ref{expbounds}) are fixed to their $d=6$
values. These bounds follow, e.g., from comparing the lower and upper
exact bounds obtained in Ref.\ \cite{WRIG86} for the noise exponent
$\kappa$ in weakly nonlinear networks, within the NLB scheme. They
are satisfied by simulation results \cite{WRIG86}. 
Using the usual
effective-medium values $s=t=1$ in (\ref{expbounds}) implies the
absurd value $\nu=0$, save for $d=2$ where a finite value of $\nu$ is
allowed. Though information about the correlation length $\xi$ (and
therefore about $\nu$) is not included in the Bruggeman nor in the HY
formulas, the above bounds show that, as long as they are meant to
model percolating systems obeying the NLB picture, these formulas are
truly adequate only in dimension $d=2$ -- and $d=1$ where the Bruggeman
formula is exact. As to ``type 2'' formulas, we insert the
values $s=0$, $t=2$ into (\ref{expbounds}) and deduce that
$\nu=1/(d-2)$, a reasonable expression for $d\geq 3$ only. If we
furthermore insist on having $\nu\geq 1/2$ as in real systems, these
heuristic arguments restrict the range of validity of the new formulas
to $d=3,4$. Note, moreover, that only in dimension $d=2$ are the
exponents equal: $s=t$, because of self-duality \cite{STRA77}.  A
formula with unequal exponents therefore is expected to be essentially
relevant to dimensions $\geq 3$.

We also quote theoretical bounds for $t$ due to Golden, valid for
hierarchical NLB models: $1\leq t\leq 2$ for $d=2,3$ and $2\leq t\leq
3$ for $d\geq 4$ \cite{GOLD90}. The above analysis is consistent with
these bounds, and can be summarized as a set of prescriptions for
using the ``best'' available effective medium theories, as far as a
non-conflicting critical behaviour is concerned: for $d=1$,
Bruggeman's formula, or its ``type 2'' counterpart are exact; for
$d=2$, the Bruggeman or HY formulas are adequate; for $d=3,4$, ``type
2'' formulas are applicable; finally, for $d\geq 5$ mean-field
theories would be the most relevant.  Estimates or exact values for
the exponents are \cite{STAU92}: $(\nu,s,t)=(4/3,1.3,1.3)_{d=2}$,
$(0.88,0.73,2.00)_{d=3}$, $(0.68,0.4,2.4)_{d=4}$,
$(0.57,0.1,2.7)_{d=5}$, $(1/2,0,3)_{d\geq 6}$. These values support
our prescriptions.

An interesting observation is that actually both ``type 1'' 
and ``type 2'' formulas can be given by a variational formulation as
\begin{eqnarray}
\label{min}
\sigma_{\text{eff}}^{\text{type 1}}&=& \min_{\sigma_0\geq 0\atop 0\leq
q(\sigma_0)\leq 1}\sigma_{\text{eff}}(\sigma_0),\\
\label{max}
\sigma_{\text{eff}}^{\text{type 2}}&=&
\max_{\sigma_0\geq 0\atop 0\leq q(\sigma_0)\leq 1}
\sigma_{\text{eff}}(\sigma_0),
\end{eqnarray}
provided that an unphysical solution $\sigma_{\text{eff}}=0$ is
discarded in the minimization (\ref{min}). Indeed, at least in the
framework of the two different models introduced in Sec.\ \ref{tm},
the curves for $q(\sigma_0)$ and $\sigma_{\text{eff}}(\sigma_0)$ are
found to have the form shown in Fig.\ 6. The infimum (\ref{min})
occurs at $q=0$, whereas the solution $\sigma_0^*$ to the equation
$\sigma_{\text{eff}}'(\sigma_0)=0$ corresponds to a maximum of
$\sigma_{\text{eff}}$. Both types of theories can therefore be
interpreted as extremal theories in the framework of self-consistent
models built on the replica-coupling ansatz.  The physical meaning of
this interpretation is still not clear. However, ``type 2''
formulas should not been disregarded as unphysical because of their
showing up as maximal ones: the minimization principle states that the
dissipated power is minimized with respect to the electric field; but
there is no reason why an extremization with respect to arbitrary
variational parameters should not lead to a maximum of the dissipated
power.  Eqs.\ (\ref{min}), (\ref{max}) explain why for a given
approximation, one always has $\sigma_{\text{eff}}^{\text{type 1}}
\leq\sigma_{\text{eff}}^{\text{type 2}}$ in Figs.\ 1-4.
 
We now consider some points that were not explicitly treated in the
paper. First, we presented the formalism in terms of the electric
field $E$, from which we obtained a conductivity
$\sigma_{\text{eff}}$. The electric current $j$ (or the induction
$D$), could be used instead \cite{BART98}. In such a formulation, the
random constitutive parameter is the resistivity $\rho(x)=1/\sigma(x)$
and the constraints are $\nabla\cdot j=0$ and $\overline{j}=j_0$. One
then computes an effective resistivity $\rho_{\text{eff}}$. Both
formulations are equivalent, but a given approximation scheme in
general leads to different results for $\sigma_{\text{eff}}$ and
$\tilde\sigma_{\text{eff}}=1/\rho_{\text{eff}}$.  Preliminary
investigations of ``type 2'' formulas have been led in this
case. These will be presented elsewhere.  Finally, we discuss the
natural question about the possibility of replica-symmetry breaking
\cite{PARI84}. Replica symmetry breaking introduces more free
parameters in the ansatz, and the final extremization has to be
carried out with respect to several variables. There is no frustration
in this problem, and we therefore expect the replica symmetric
solution to be the only one. In order to test this, we tried a one-step
symmetry-breaking solution and did indeed not find any new solution.

\section{Conclusion}
\label{c}

We presented a functional approach to the calculation of
effective-medium properties of random media. We showed how to recover
the Bruggeman and Hori-Yonezawa formulas by using specific
approximation schemes to the basic functional integral.  We also
discussed the introduction of a replica-coupling parameter in a
gaussian Ansatz, from which new effective-medium formulas were
obtained. These formulas appear to be more adequate in $d=3$ compared
to the standard ones by Bruggeman and HY. Because it yields a
sensible result in all dimensions, and fulfills all the constraints
required to deserve the label of a ``good'' effective-medium theory,
the ``type 2'' counterpart of the Bruggeman formula offers an
interesting alternative to the latter. Indeed, it has a percolation
threshold equal to $p_c=1/5$ in three dimensions. This is closer to
values observed in real materials, compared to the $p_c=1/3$ of the
Bruggeman formula which often constitutes an overestimation.

\acknowledgements

We gratefully acknowledge H.~Orland for stimulating discussions. One
of us (MB) wants to thank H.E.~Stanley for his hospitality at the CPS
and the DGA for financial support.

\appendix
\section{Quadratic fluctuations of the field}
\label{qfotf}
For completeness, we give here the demonstration of Eq.\ (\ref{berg})
\cite{BERG78}. Volume averages are identified to statistical ones.
Because $W^*=(1/2)\langle \sigma E^2\rangle =(1/2)\sum_\nu p_\nu
\sigma_\nu \langle E^2\rangle_\nu$, where $\langle \cdot \rangle_\nu$
denotes an average on the phase $\nu$ on which the conductivity
$\sigma_\nu$ is constant, the effective conductivity reads
\begin{equation}
\sigma_{\text{eff}}=\sum_\nu p_\nu \sigma_\nu \frac{\langle E^2\rangle_\nu}{E_0^2}.
\end{equation}
On the other hand, $\sigma_{\text{eff}}$ has to be an homogeneous 
function of degree one of the $\sigma_\nu$, whence
\begin{equation}
\sigma_{\text{eff}}=\sum_\nu \sigma_\nu \frac{\partial \sigma_{\text{eff}}}{\partial \sigma_\nu}.
\end{equation}
Comparing both equations yields the values of the 
$\langle E^2\rangle_\nu$ and consequently that of $\langle E^2\rangle
=\sum_\nu p_\nu \langle E^2\rangle_\nu$. Equ.\ (\ref{berg}) follows. 
We note that if $\sigma_{\text{eff}}=\langle\sigma\rangle$ (an exact 
upper bound for the effective conductivity), then 
$\langle \Delta E^2\rangle=0$. Therefore, $\sigma_{\text{eff}}=\langle\sigma\rangle$ defines a trivial model 
of a medium which is a composite, but from which field fluctations 
are nonetheless absent. 

\section{Calculations in the ``one-impurity'' approximation}
\label{appb}
The approximation which leads to (\ref{oneimp}) is built as follows. We first expand the exponential
\begin{equation}
\label{ser}
\left\langle e^{-\beta({\cal H}_e-{\cal H}_0)}\right\rangle_0=\sum_{k\geq 0}
{(-\beta)^k\over k!} \left\langle({\cal H}_e-{\cal H}_0)^k\right\rangle_0.
\end{equation}
Since ${\cal H}_0$ is non-random, using the hamiltonian 
density $w_0$ defined in (\ref{hdens}) and $\Delta w_x[E(x)]$ defined by (\ref{dwx}) we can rewrite  the difference ${\cal H}_e-{\cal H}_0$ as
\begin{equation}
{\cal H}_e-{\cal H}_0=\int dx\, \Delta H(x),
\end{equation}
where 
\begin{equation}
\label{dh}
\Delta H(x)=-\frac{1}{\beta v}\ln\left\langle e^{-\beta v \Delta w_x[E(x)]}\right\rangle.
\end{equation}
The one-impurity approximation consists in writing ($k\geq 1$)
\begin{mathletters}
\begin{eqnarray}
\label{intint}
({\cal H}_e-{\cal H}_0)^k
&=&
\int dx_1\ldots dx_k\,\Delta H(x_1)
\ldots
\Delta H(x_k)\\ 
&\simeq& v^{k-1}\int dy\, 
\Delta H(y)^k.
\end{eqnarray}
\end{mathletters}
The last expression only retains contributions from identical points 
in Eq.\ (\ref{intint}). Summing back the series in (\ref{ser}), and using (\ref{dh}) yields
\begin{eqnarray}
\label{ebracko}
&&\left\langle e^{-\beta({\cal H}_e-{\cal H}_0)}\right\rangle_0
\simeq 
1+\int {dy\over v}
\left\langle {\cal I}(y)-1\right\rangle,\\
\label{caliy}
&&{\cal I}(y)=\frac{\int \tilde{\cal D}(E^{\alpha})\,
e^{-\beta\int dx\, \left\{w_0[E(x)]
+v\,\Delta w_{\bf x}[E(x)]
\delta(x-y)\right\}}}
{\int \tilde{\cal D}(E^{\alpha})\,e^{-\beta {\cal H}_0}},
\end{eqnarray}
where a one-impurity-type integral is involved. Since the fundamental
size of the theory ($\sim v^{1/d}$) is much smaller than the volume
$V=1$ of the system, and since the latter is statistically
translation-invariant, the outer integral over $y$ is redundant with
the disorder average, and can be dropped. We therefore arrive at
(\ref{oneimp}).

When $\beta\to\infty$, the functional ${\cal I}(y)$ can be computed
exactly. Let us briefly indicate how to do it. We first introduce the
notation $\vec{h}$ for vectors of dimension $nd$, and components
$h^\alpha_i$, with $\alpha=1,\ldots,n$, $i=1,\ldots,d$.  Hence,
$\Delta w_y[E(y)]\equiv \Delta w_y\bigl(\vec{E}(y)\bigr)$. The next
step is to use the formal identity
\begin{equation}
e^{-\beta v\Delta w_y(\vec{E})}=\int {d\vec{h}\,d\vec{h}'\over
(2\pi)^{nd}}e^{-i\vec{h}\cdot \vec{h}'} e^{-\beta v\Delta
w_y\bigl(-i\frac{\partial}{\partial \vec{h}'}\bigr)} e^{i
\vec{h}'\cdot \vec{E}}
\end{equation}
to write the numerator of (\ref{caliy}), which we denote hereafter by
${\cal J}(y)$, as
\begin{equation}
{\cal J}(y)=\int {d\vec{h}\, d\vec{h}'\over
(2\pi)^{nd}}e^{-i\vec{h}\cdot \vec{h}'} e^{-\beta v \Delta
w_y\bigl(-i\frac{\partial}{\partial\vec{h}'}\bigr)}\int\tilde{\cal
D}E\,e^{-(\beta/2)\int dx\, \vec{E}(x)\cdot \tilde{M} \cdot
\vec{E}(x)+i\vec{h}'\cdot \vec{E}(y)},
\end{equation}
where $\tilde{M}$ is the matrix defined from the replica-coupling
matrix $M$ in $w_0$ by $\tilde{M}^{\alpha\gamma}_{ij}\equiv
M^{\alpha\gamma}\delta_{ij}$. After an integration over the fields $E$
and $\phi$ implied in the measure $\tilde{\cal D}E$ (which can be
easily done using the Fourier components of $\phi$, and with $y=0$
since (\ref{oneimp}) is independent of $y$), $\cal J$ reads, up to
inessential factors \cite{NOTE2}:
\begin{equation}
{\cal J}(y)=e^{-(\beta/2)\sum_{\alpha\gamma}M^{\alpha\gamma}E_0^2}
\int {d\vec{h}\, d\vec{h}'\over (2\pi)^{nd}}e^{-i\vec{h}\cdot \vec{h}'}
e^{-v\beta\Delta w_y\bigl(-i\frac{\partial}{\partial \vec{h}'}\bigr)}
e^{i \vec{h}'\cdot \vec{E}_0-\vec{h}'\cdot \tilde{M}^{-1}
\cdot \vec{h}'/(2\beta v d)}.
\end{equation}
Formally expanding $\exp(-v\beta\Delta w_y)$ in powers of 
$-i\partial/\partial \vec{h}'$, and carrying out successive 
integrations by parts over $\vec{h}'$ yields
\begin{equation}
{\cal J}(y)=\left[\mathop{\text{Det}}(M)\left(\frac{v\beta
d}{2\pi}\right)^n\right]^{d/2}
e^{-(\beta/2)\sum_{\alpha\gamma}M^{\alpha\gamma}E_0^2}\int d\vec{h}\,
e^{-(v \beta d/2)(\vec{E}_0-\vec{h})\cdot
\tilde{M}\cdot(\vec{E}_0-\vec{h})-v\beta\Delta w_y(\vec{h})},
\end{equation}
where the determinant is evaluated in replica space. Finally, 
${\cal I}(y)={\cal J}(y)/{\cal J}(y;\Delta w_y=0)$:
\begin{equation}
\label{iy}
{\cal I}(y)=\left[\mathop{\text{Det}}(M)\left(\frac{v\beta
d}{2\pi}\right)^n\right]^{d/2}\int \prod_\alpha dh^\alpha \,
e^{-v\beta\left\{{d\over 2}
\sum_{\alpha\gamma}M^{\alpha\gamma}(E_0-h^\alpha)_i
(E_0-h^\gamma)_i+\Delta w_y[h]\right\}}.
\end{equation}
For any $\Delta w_y$, this integral over the replicated vector field
$h$ can be computed exactly using a saddle-point method \cite{NEGE87}
in the limit $\beta\to\infty$, as announced. This allows for a
possible extension of the theory to nonlinear media in the
``one-impurity'' approximation. Here, for the linear problem at hand,
Eq.\ (\ref{iy}) is a simple gaussian integral. Setting
$\Delta\sigma=\sigma-\sigma_0$ and
\begin{equation}
\mu=\left(1+\frac{\Delta\sigma}{d\sigma_0}\right)^{-1},
\end{equation}
we obtain
\begin{equation}
\ln{\cal I}(y)=\left[-\frac{\beta v}{2}\Delta\sigma\,\mu E_0^2
+\frac{d}{2}(\ln \mu-\beta v q \sigma\mu E_0^2/d)\right]n+O(n^2),
\end{equation}
from which follows (\ref{dwoi}).

\section{Calculation in the ``cumulant series'' approximation}
\label{co}
We have to compute $\left\langle {\cal H}_e\right\rangle_0$ and 
$\left\langle {\cal H}_0\right\rangle_0$ in Equ.\ (\ref{dwsa}).
The calculation of $\left\langle {\cal H}_0\right\rangle_0$ is easy
with the methods already employed, and yields
\begin{equation}
\lim_{n\to 0}\left\langle 
{\cal H}_0\right\rangle_0/n =\frac{1}{2}\sigma_0 E_0^2.
\end{equation}
As to $\left\langle {\cal H}_e\right\rangle_0$, 
we first expand ${\cal H}_e$ (Eq.\ (\ref{calf})) in the cumulants
$C_k$ of the disorder averages of $\sigma(x)$, according to their
definition by the generating funtion ($X$ is a generic expansion
variable)
\begin{equation}
\ln\left<e^{X\sigma}\right>=\sum_{k\ge 1} {X^k\over k!}C_k(\sigma).
\end{equation}
We therefore have:
\begin{equation}
\label{fexpand}
{\cal H}_e=-{1\over \beta}\sum_{\bf x}\sum_{k\ge 1}{1\over k!}
\left[-{\beta v\over 2}\sum_\alpha {E^\alpha}({\bf x})^2\right]^k
C_k(\sigma).
\end{equation}
We deduce that
\begin{equation}
\label{fob}
{1\over n}\left\langle{\cal H}_e\right\rangle_0=-{1\over V\beta}
\sum_{\bf x}\sum_{k\ge
1}{(-\beta v)^k\over k!}  C_k(\sigma)\,{\cal C}_k(E^2/2),
\end{equation}
where
\begin{equation}
\label{calck}
{\cal C}_k(E^2/2)={1\over n} 
\left\langle \left[\sum_\alpha {E^\alpha(x)}^2/2\right]^k
\right\rangle_0
\end{equation}
(because of statistical homogeneity, these coefficients do not depend
on the position variable $x$). It is convenient to introduce the
following generating function ${\cal Z}(X)$ in order to compute the ${\cal C}_k$:
\begin{eqnarray}
{\cal Z}(X)&=&\sum_{k\geq 1} {(-X)^k\over k!}{\cal C}_k(E^2/2)
\nonumber\\
&=&{1\over n} \left\{\left\langle\exp\left[-{1\over 2}X 
\sum_\alpha{E^\alpha}^2(x)\right]
\right\rangle_0-1\right\}
\nonumber\\ 
&=&{1\over n}\ln\left\langle\exp
\left[-{1\over 2}X \sum_\alpha{E^\alpha}^2(x)\right]
\right\rangle_0+O(n).
\end{eqnarray}
Setting $A^{\alpha\gamma}=\delta_{\alpha\gamma}+(X/v\beta d) [M^{-1}]^{\alpha\gamma}$, we obtain \cite{NOTE2}
\begin{equation}
\label{zx}
{\cal Z}(X)=-{1\over 2n} \left\{d\mathop{\text{Tr}}
\mathop{\text{Ln}} A+X E_0^2
\sum_{\alpha\gamma} [A^{-1}]^{\alpha\gamma}\right\}+O(n)
\end{equation}
(the trace and the logarithm act in the replica space).

Expanding (\ref{zx}) in powers of $X$ then allows for the identification
\begin{equation}
{\cal C}_k(E^2/2)={k!\over 2}\left({1\over v\beta d}\right)^k \frac{d}{n}\left[{1\over k}\mathop{\text{Tr}} (M^{-k})+{v\beta d} E_0^2 \sum_{\alpha\gamma} [M^{1-k}]^{\alpha\gamma}\right]+O(n).
\end{equation}
Use of this expression in (\ref{fob}) cancels the convergence factor
$k!$: we reintroduce it by inserting the identity
\begin{equation}
{1\over m!}\int_0^\infty du\, e^{-u} u^m =1,
\end{equation}
applied to $m=k-1$ and $m=k$ in the resulting cumulant series. This
permits its Borel summation, which brings in the functions $h_m(x)$
defined by (\ref{hm}). This results in 
\begin{equation}
\label{fmfo}
{1\over n}
\langle{\cal H}_e\rangle_0
=-{1\over n}
\left[ 
{d\over 2}E_0^2
\sum_{\alpha\gamma}\left[ M
h_0(M^{-1}/d)\right]^{\alpha\gamma}
+{1\over 2 v \beta}\mathop{\text{Tr}}\, h_{-1}(M^{-1}/d)
\right]+O(n),
\end{equation}
where we defined the family of functions
\begin{equation}
\label{hm}
h_m(z)=\int_0^\infty du\,u^m e^{-u} \ln\left\langle e^{-u\sigma
z}\right\rangle\qquad(m>-2).
\end{equation}
Note for further use that
\begin{equation}
\label{deriv}
h_m'(z)={1\over z}\left[h_{m+1}(z)-(k+1)h_{m}(z)\right].
\end{equation}

For $M$ given by (\ref{matmok}), the differents terms in (\ref{fmfo}) are
\begin{eqnarray}
&&\lim_{n\to 0}{1\over n} \sum_{\alpha\gamma} \left[M
h_0(M^{-1}/d)\right]^{\alpha\gamma}=\sigma_0 h_0
\bigl(1/(d\sigma_0)\bigr),
\nonumber\\ 
&&\lim_{n\to 0}{1\over n}\mathop{\text{Tr}}
h_{-1}(M^{-1}/d)=dh_{-1}\bigl(1/(d\sigma_0)\bigr)+v\beta\, d\sigma_0 q\, h_0 \bigl(1/(d\sigma_0)\bigr)E_0^2.
\end{eqnarray}
which leads to Eq.\ (\ref{fmfodeu}).


%
\twocolumn

\begin{figure}
\narrowtext
\vspace*{0.0cm}
\centerline{
\epsfysize=0.9\columnwidth{\rotate[r]{\epsfbox{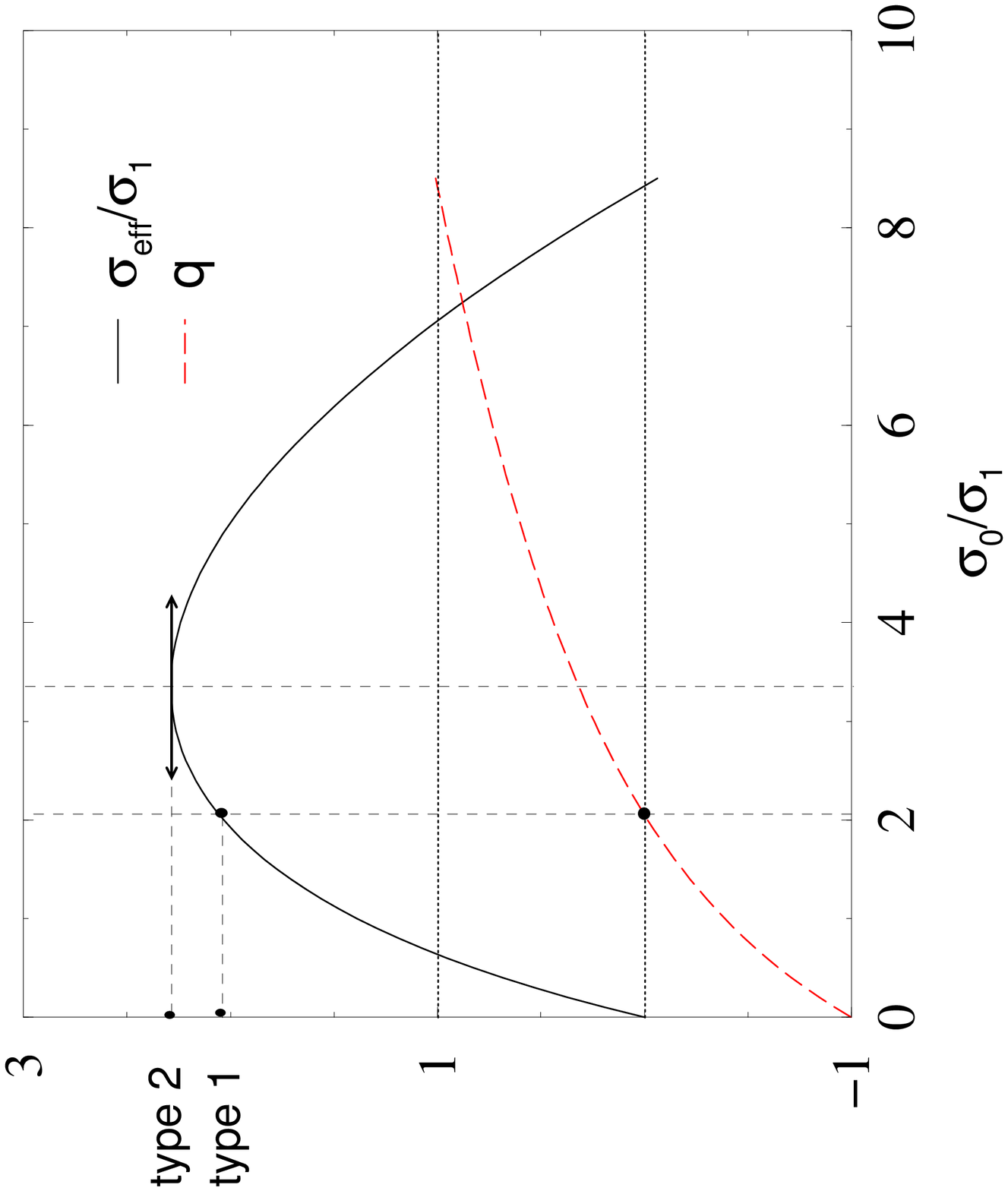}}}}
\vspace*{0.3cm}
\caption{Rescaled effective conductivity 
$\sigma_{\text{eff}}(\sigma_0)/\sigma_1$, and reduced
replica coupling parameter $q(\sigma_0)$ vs.\ $\sigma_0$ in
dimension $d=3$ for a binary medium. In this exemple computed 
from Eqs.\ (\ref{sigefq}), (\ref{qbrug}), the conductivity ratio is 
$\sigma_2/\sigma_1=100$, and the volume fraction $p$ of 
component 2 is $p=0.18$. The ``type 1'' effective conductivity is
obtained when $q=0$, whereas the ``type 2'' effective conductivity
corresponds to the maximum of the curve $\sigma_{\text{eff}}(\sigma_0)$.
}
\label{fig6}
\end{figure}


\begin{references}
\bibitem{NOTE1} This is because the key ingredient of the continuum 
theory, namely the Green function $\sum_k k_i k_j/k^2$ (computed at $x=0$ 
in the calculations of this paper, cf.\ \cite{NOTE2}), 
has the same low-momentum structure as its counterpart on an 
hypercubic lattice (see, e.g.\ Ref.\ \cite{LUCK91}).
\bibitem{NOTE2} Integration is performed over the independant scalars 
$\phi_k$ and $\phi_{-k}$ with $1/(2v)$ such couples of Fourier modes 
involved. Use is made in the calculations of the 
equality $1/v=\sum_k 1\simeq\int dk /(2\pi)^d$, with an 
upper cut-off on the $k$ implied. In particular, one appeals 
to the formula $\sum_k k_i k_j/k^2=\delta_{ij}/(dv)$.
\end{references}
\end{document}